\documentclass[floatfix,nofootinbib,twocolumn,showpacs,superscriptaddress, reprintnumbers,amssymb,letterpaper,amsmath,prl]{revtex4-2}

\usepackage{graphicx}
\usepackage{bm}
\usepackage[colorlinks=true,allcolors=blue]{hyperref}
\usepackage{lineno}
\usepackage{siunitx}
\usepackage{physics}
\usepackage{tikz}

\begin{document}


\title{Demonstration of sub-GV/m Accelerating Field in a Photoemission Electron Gun Powered by Nanosecond $X$-Band Radiofrequency Pulses}
\author{W. H. Tan}%
\affiliation{Northern Illinois Center for Accelerator \& Detector Development and Department of Physics, Northern Illinois University, DeKalb, Illinois 60115, USA}
\author{S. Antipov}%
\affiliation{Euclid Techlabs LLC, Bolingbrook, Illinois 60440, USA}%
\author{D. S. Doran}%
\affiliation{Argonne National Laboratory, Lemont, Illinois 60439, USA}
\author{G. Ha}%
\affiliation{Argonne National Laboratory, Lemont, Illinois 60439, USA}
\author{C. Jing}%
\email{c.jing@euclidtechlabs.com}
\affiliation{Euclid Techlabs LLC, Bolingbrook, Illinois 60440, USA}%
\affiliation{Argonne National Laboratory, Lemont, Illinois 60439, USA}
\author{E. Knight}
\affiliation{Euclid Techlabs LLC, Bolingbrook, Illinois 60440, USA}%
\author{S. Kuzikov}
\email{s.kuzikov@euclidtechlabs.com}
\author{W. Liu}%
\affiliation{Argonne National Laboratory, Lemont, Illinois 60439, USA}
\author{X. Lu}
\affiliation{Northern Illinois Center for Accelerator \& Detector Development and Department of Physics, Northern Illinois University, DeKalb, Illinois 60115, USA}
\affiliation{Argonne National Laboratory, Lemont, Illinois 60439, USA}
\author{P. Piot}
\email{ppiot@niu.edu}
\affiliation{Northern Illinois Center for Accelerator \& Detector Development and Department of Physics, Northern Illinois University, DeKalb, Illinois 60115, USA}
\affiliation{Argonne National Laboratory, Lemont, Illinois 60439, USA}
\author{J. G. Power}%
\affiliation{Argonne National Laboratory, Lemont, Illinois 60439, USA}
\author{J. Shao}%
\affiliation{Argonne National Laboratory, Lemont, Illinois 60439, USA}
\author{C. Whiteford}%
\affiliation{Argonne National Laboratory, Lemont, Illinois 60439, USA}
\author{E. E. Wisniewski}%
\affiliation{Argonne National Laboratory, Lemont, Illinois 60439, USA}

\date{\today}

\begin{abstract}

Radiofrequency (RF) electron guns operating at high accelerating gradients offer a pathway to producing bright electron bunches. Such beams are expected to revolutionize many areas of science: they could form the backbone of next-generation compact x-ray free-electron lasers or provide coherent ultrafast quantum electron probes. We report on the experimental demonstration of an RF photoemission electron source supporting an accelerating field close to 400~MV/m at the photocathode surface. The gun  was operated in an RF transient mode driven by short $\sim 9$~ns X-band (\SI{11.7}{\giga\hertz}) RF pulses. We did not observe any major RF breakdown, or significant dark current over a three-week experimental run at high accelerating fields. The demonstrated paradigm provides a viable path to  forming relativistic electron beams with unprecedented brightness. 
\end{abstract}

\maketitle

Charged-particle accelerators have been invaluable engines of discovery in fundamental sciences since their inception in the early 20's century.  Bright electron beams have especially enabled the development of accelerator-based light sources, i.e. free-electron lasers (FELs), capable of producing coherent radiation over a broad range of the electromagnetic spectrum~\cite{bonifacio-1984-a}. Additionally, directly employing bright electron beams as primary probes, e.g., in ultra-fast electron scattering setups, has produced groundbreaking research in condensed matter and chemistry~\cite{dwyer-2006-a}. A critical aspect to the generation of bright electron beams is the rapid acceleration of the bunch during the emission process to mitigate space-charge effects. Specifically, the beam peak brightness ${\cal B}=q/\Gamma$, where $q$ is the bunch charge and $\Gamma$ its six-dimensional phase-space volume, scales with the electric field $E_a$  experienced during emission at the emitter surface as ${\cal B} \propto E_a^{\alpha}$  where the exponent $\alpha\ge 1$ depends on the bunch's initial transverse-to-longitudinal aspect ratio~\cite{bazarov-2009-a,filippetto-2014-a}.

Normal-conducting radiofrequency (RF) electron sources~\cite{fraser-1986-a,mcdonald-1988-a} (``RF guns") invented in the mid 80’s support high electric fields which ultimately boosted the development of X-ray FELs~\cite{mcneil-2010-a}. The highest field an RF resonator can sustain is ultimately limited by the peak surface field $E_0$  on its walls. The performances of the high-gradient electron sources ($E_0>200$~MV/m) designed in the recent years~\cite{limborg-2016-a,marsh-2018-a,othman-2020-a} have been hindered by RF breakdown and pulse heating -- phenomena where the surface of the resonator or emitter deteriorates thus momentarily preventing the storage of electromagnetic energy in the resonator~\cite{fursey-1985-a}. Besides, these sources often exhibit substantial ``dark-current" emissions due to the spurious uncontrolled emission of electrons via quantum tunneling from the surfaces exposed to the high electric fields~\cite{fowler-1928-a}. 
In this letter, we demonstrate that the use of short-pulse electromagnetic fields offers a viable path toward the generation of \si{\giga\volt/\metre} surface fields in metallic accelerating structures. We specifically confirm the attainment of $>$\SI{0.5}{\giga\volt/\metre} electric field on the copper surface of a resonant cavity operating at \SI{11.7}{\giga\hertz} and powered by 9-ns RF pulses. The surface field is calculated by measuring the energy of the electron bunches photoemitted by impinging an ultraviolet laser pulse on the surface. The required high-peak-power short RF pulses are produced via deceleration of a train of relativistic electron bunches ($\sim$\SI{60}{\mega\electronvolt}) in a slow-wave structure. %

Over the past two decades, experimental investigations have shown that the RF-breakdown process in structures operating at room temperature is predominantly correlated with the surface electric field, RF-pulse width, and pulse-heating temperature~\cite{Dolgashev-2010-a,Laurent-2011-a,grudiev-2009-a}. A phenomenological model based on experiments conducted with RF-pulse duration $\tau_p \sim[100,1000]$~ns suggests that the breakdown rate (BDR) follows a power law of the form $
\mbox{BDR} \propto E_0^{30} \tau_p^5,$~\cite{grudiev-2009-a,doebert-2007-a,wang-2009-a}. The latter equation implies that, for a given BDR, an increase in the field amplitude $E_0$ can be accommodated by shortening the RF-pulse duration. This approach is the basis of the present work. It should be noted that recent experiments also point to an alternative path to lowering the BDR using cryogenic-cooled copper resonators~\cite{cahill-2018-b}. Another benefit of the short RF pulses is the mitigation of the average dark-current, i.e., the total charge emitted from the structure via mechanisms other than photoemission  (i.e. field emission). In our case, the acceleration field only lasts for 9~ns so that the average dark current is orders of magnitude smaller than in conventional guns powered by \textmu{s} RF pulses. 

\begin{figure}[hhhh!]
\includegraphics[width=0.49\columnwidth]{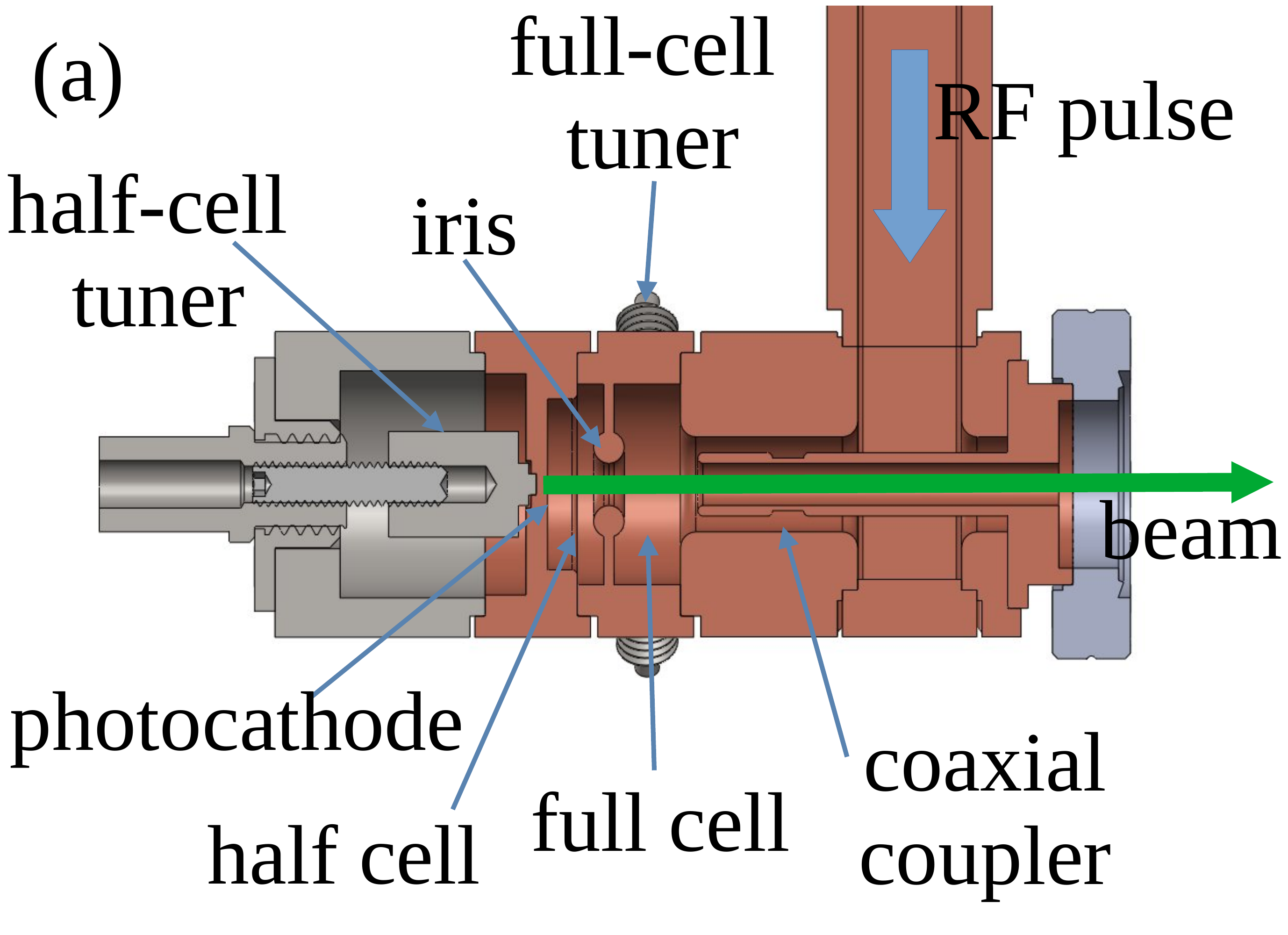}
\includegraphics[width=0.49\columnwidth]{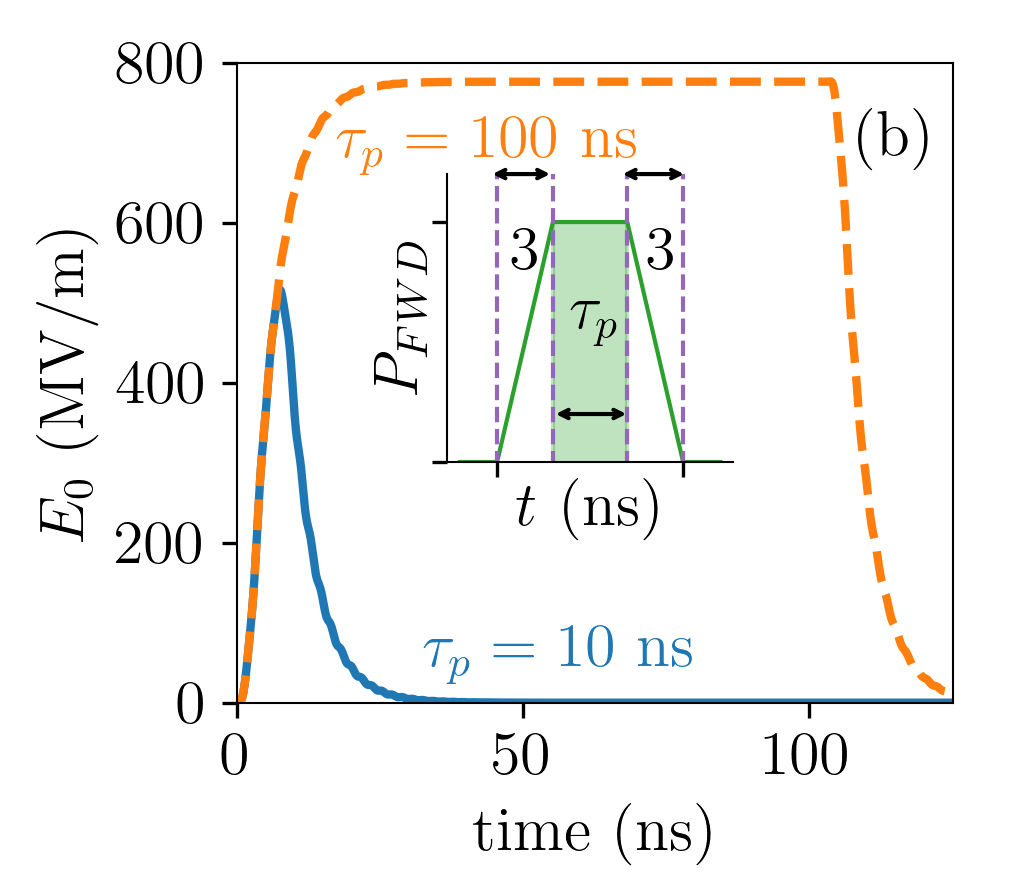}
\centerline{\includegraphics[width=0.5\textwidth]{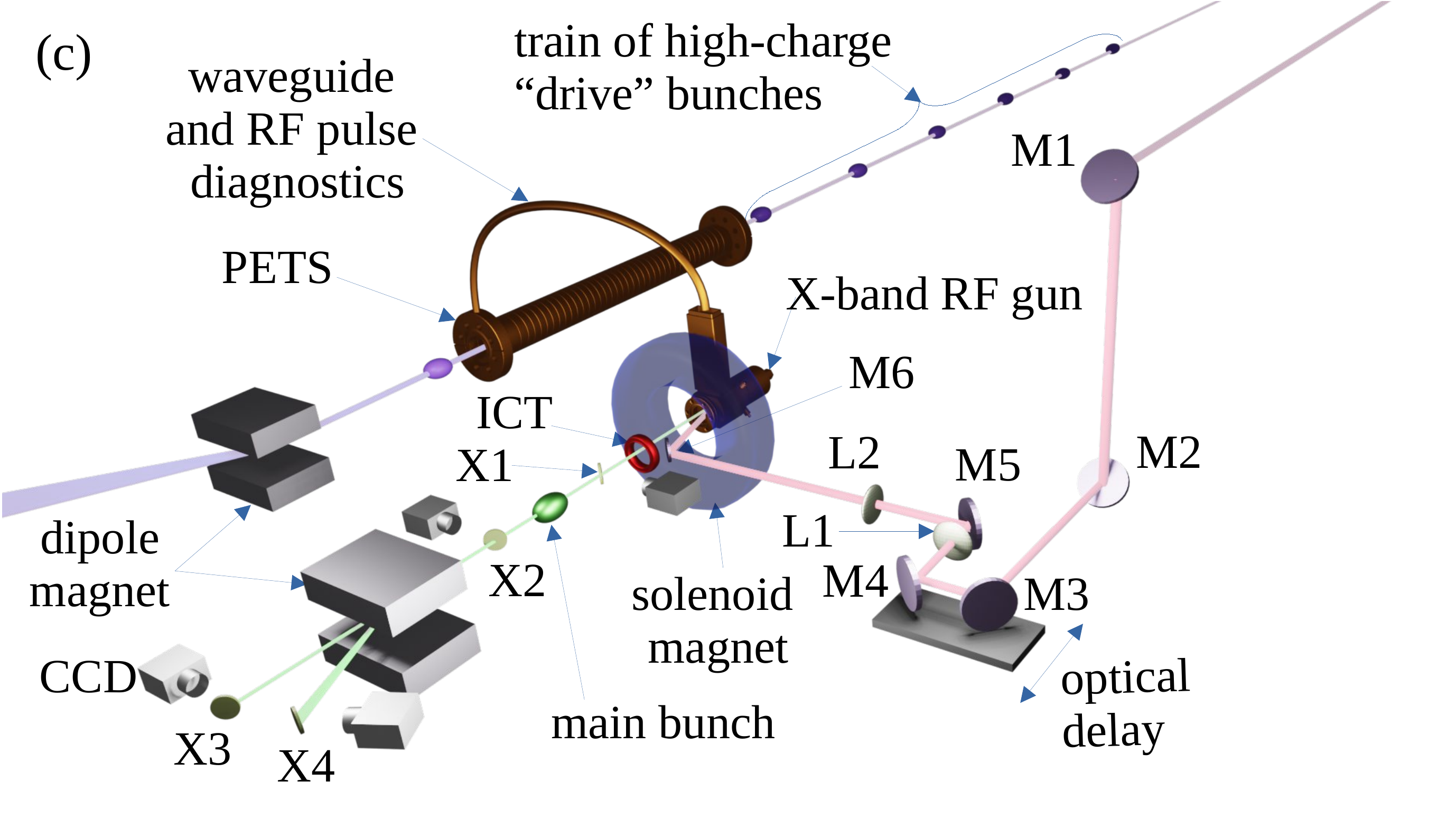}}
\caption{Schematic of the  XRF gun (a), and simulated peak field on photocathode (b) for a short ($\tau_p=3$-ns, solid trace) and long ($\tau_p=100$-ns, dash trace) flat-top 300-MW peak power RF pulse (the inset details the forward-power $P_{FWD}$ RF-pulse profile with 3-ns rising and falling edges). Overview of the experimental setup (a): the upper (magenta) path corresponds to the relevant section of the AWA main accelerator, the lower (green) path is the XRF gun beamline. The pink path shows the UV laser path. The labels X$i$, M$i$, and L$i$ respectively refer to Ce:YAG scintillating screens with CCD cameras, optical mirrors, and lenses. \label{fig:expconfig}}
\end{figure}
To investigate the generation of transient high electric fields using high-peak-power short RF pulses, an X-band RF (XRF) photoemission gun was designed~\cite{kuzikov-2019-a}. The gun consists of a $1+\frac{1}{2}$-cell resonator functioning on the TM$_{010,\pi}$ mode at 11.7~GHz; see Fig.~\ref{fig:expconfig}(a). It is a strongly over-coupled resonator resulting in a low loaded quality factor ($Q_{\ell}\simeq 180$), thus rapidly establishing the RF field inside the cavity. The gun is made of oxygen-free copper using conventional fabrication techniques combining high-precision machining and hydrogen-furnace brazing. A broadband coaxial RF-input coupler is used to ensure the field distribution remains axisymmetric and ultimately maximize the beam brightness. The iris disk located between the two cells incorporates four magnetic coupling slots which increase both the RF coupling between cells and the frequency separation from the neighboring resonant TM$_{010,0}$ mode. Figure~\ref{fig:expconfig}(b) compares the simulated peak field on the photocathode for the XRF gun being driven by a long ($\tau_p$=100~ns) and a short ($\tau_p$=3~ns) RF pulses and reveals that the maximum field amplitude attained in the transient regime is 67\% of the steady-state field (i.e. reached after time $t\simeq 30$~ns). In summary, the advantage of operating the gun in the short-pulse mode, as opposed to the long-pulse mode, is that the photocathode experiences the field for 1/10th the time. This, in turn, leads to the ability to operate the gun at higher gradient with simultaneously lower dark current.

The XRF gun was integrated into the Argonne Wakefield Accelerator (AWA)~\cite{conde-2017-a}. The AWA includes an L-band ($f_L=1.3$~GHz) RF photoinjector that produces an electron beam arranged as a train of 8 high-charge electron bunches separated by $T_L=1/f_L\simeq 769$~ps. The bunch train (with total charge $Q_{db}\le 400$~nC) is further accelerated in a linear accelerator to a final energy of $\sim \SI{60}{\mega\electronvolt}$ before being transversely focused in a power-extraction and transfer structure (PETS) to excite high-peak-power RF pulse at $f=9f_L=11.7$~GHz~\cite{shao-2019-a}. For the experiment reported below, the RF-pulse power was $\sim 250$~MW (and occasionally up to 300~MW) with a typical RF pulse shape displayed in Fig.~\ref{fig:expconfig}(b, inset). The RF pulses were out-coupled from the PETS and guided to the XRF-gun collocated to the AWA accelerator; see Fig.~\ref{fig:expconfig}(c). Owing to the 2-Hz repetition rate of the AWA facility, only a few Watts of average RF power is dissipated in the XRF gun despite operating in such a high-peak-power regime, thus no water cooling is required.

The XRF gun is nested in a magnetic solenoidal lens that provides control over the beam transverse size and eventually maximizes the beam brightness~\cite{serafini-1997-a,tan-2021-a}. 
The downstream beamline includes remotely-insertable Ce:YAG scintillating screens to measure the transverse beam distribution, and one integrated current monitor (ICT) for bunch-charge measurement. One of the screens (X4) is located at a high-dispersion point downstream of a dipole magnet for energy measurement. The bunches are generated by impinging an ultraviolet (UV) laser pulse ($\le 200$~\textmu{J}) split off from the $\sim 4$-mJ laser pulse producing the high-charge drive bunches. The low-intensity pulse is sent to the experimental area (from mirror M1) after propagating in a $\sim 30$-m optical line with imaging lenses and a coarse delay that ensures the laser pulse impinges the photocathode within $\sim 2$~ns of the RF-pulse injection in the XRF gun. In the experimental area, the laser path comprises a fine adjustable optical delay line (M3-M4) with range $\pm 2$~ns, along with a set of lenses (L1, L2) to focus the beam on the photocathode. The laser enters the evacuated XRF-gun beamline where an in-vacuum mirror (M6) directs the pulse on the back-plane of the XRF gun which serves as a copper photocathode; see  Fig.~\ref{fig:expconfig}(a). The UV laser on the photocathode has a uniform distribution with a radius of 400~\textmu{m} with a $\sim 400$-fs duration. 

\begin{figure}[ttttt!!!!!!]
   \centering
   \includegraphics[width=0.95\columnwidth]{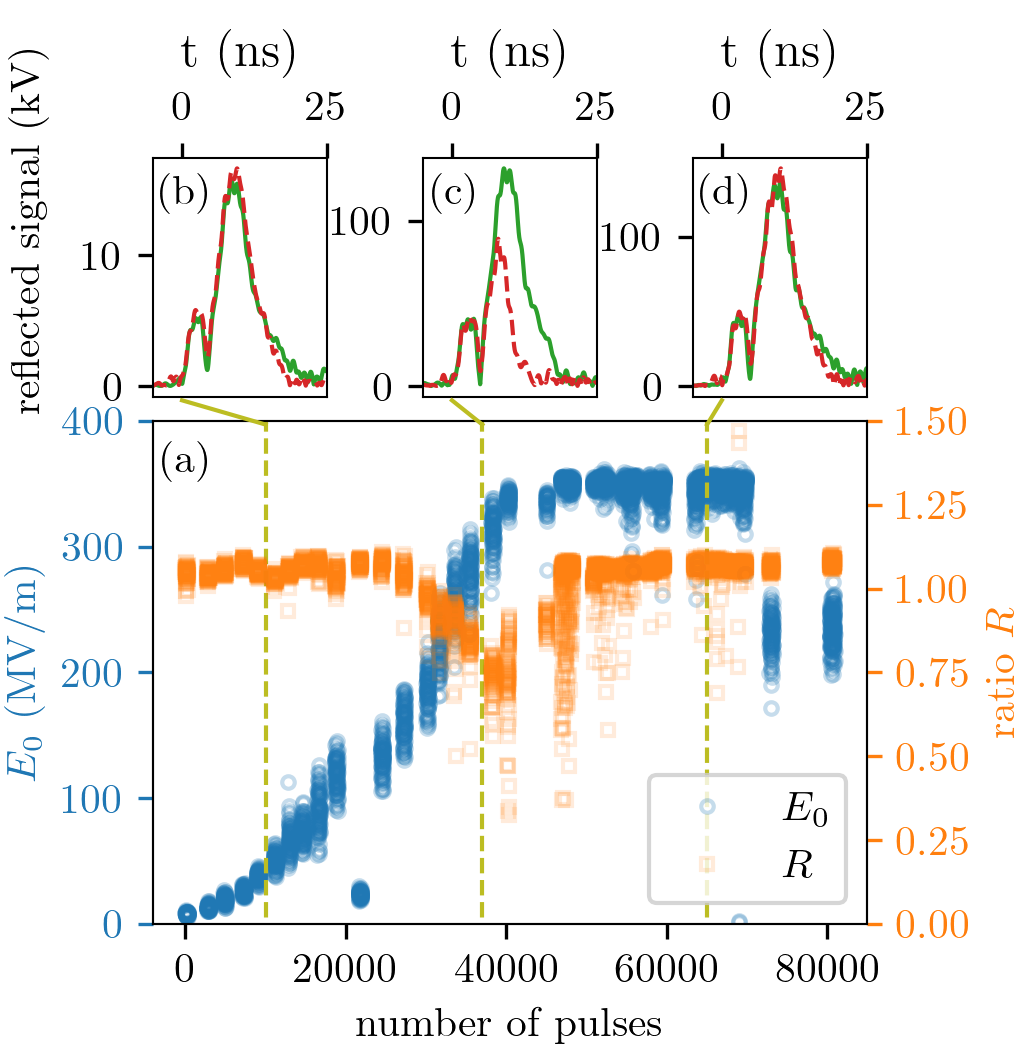}
\caption{RF conditioning history of the XRF gun (a) where the photocathode peak field $E_0$ was gradually increased (blue circle) and the ratio of peak value of the  measured to simulated reflected RF signals, $R$, recorded (orange square). The upper plots (b-d) compared the measured (dash red trace) and simulated (green solid trace) rectified RF-reflection signals at different stage of the conditioning history indicated with the dash vertical lines.}
   \label{fig:rfconditioning}
\end{figure}

The XRF gun was conditioned to photocathode fields $E_0\simeq 350$~MV/m deduced from the measured forward RF power. Figure~\ref{fig:rfconditioning}(a) shows the RF-conditioning history with the evolution of the field amplitude as a function of the number of RF pulses. The latter figure also reports the ratio of the measured over simulated peak values of the reflected-power signal. A value of $R=1$ implies there are no RF breakdowns (or dark-current-induced beam loading). Due to the non-ideal charge balance of the drive bunch train, the forward-power flat-top duration was $\tau_p< 3$~ns. In the simulation, we used the measured charge balance combined with the frequency-domain bench measurements of the XRF gun to calculate the input and reflected RF pulse shapes.
From Fig.~\ref{fig:rfconditioning}(a), we observe three conditioning regions. First, for values $E_0\le $~\SI{150}{\mega\volt/\metre} the field increases monotonically with $R\simeq 1$. Then, at $E_0\simeq$~ \SI{150}{\mega\volt/\metre}, the ratio $R\le 1$ and decreases until it reaches its minimum for fields  $E_0\simeq$~\SI{250}{\mega\volt/\metre} due to breakdown event and dark-current emission. Eventually, the XRF gun is conditioned and a value $R\simeq 1$ is recovered for $E_0\simeq 350$~MV/m (after $5\times 10^4$ RF pulses, with forward power $P_{FWD} \simeq 250$~MW corresponding to $Q_{db}\simeq 350$~nC). At the time of the conditioning, the maximum field amplitude attained was capped by the available RF power. The entire conditioning process took only $7\times10^4$ pulses (i.e. $\sim 10$ hours at \SI{2}{\hertz} repetition rate). Ultimately, the field was set to $E_0 \simeq$ \SI{200}{\mega\volt/\metre}, to confirm the absence of any additional breakdown events or dark-current emission. Subsequently, the XRF gun was operated up to \SI{\sim 400}{\mega\volt/\metre} during the photoemission experiment and no dark current could be detected on the ICT or Faraday cup. We resorted to using the pixels noise level associated with the data recorded at X1 while scanning the solenoid magnetic field to estimate an upper bound value of $\sim 1$~pC for the dark-current charge emitted over the entire RF pulse. This estimate assumes the number of electrons is proportional to the light intensity in each pixel which was a priori calibrated against the ICT for the photoemitted beam. 

\begin{figure}[bbbbb!!!!!]
   \centering
   \includegraphics[width=0.915\columnwidth]{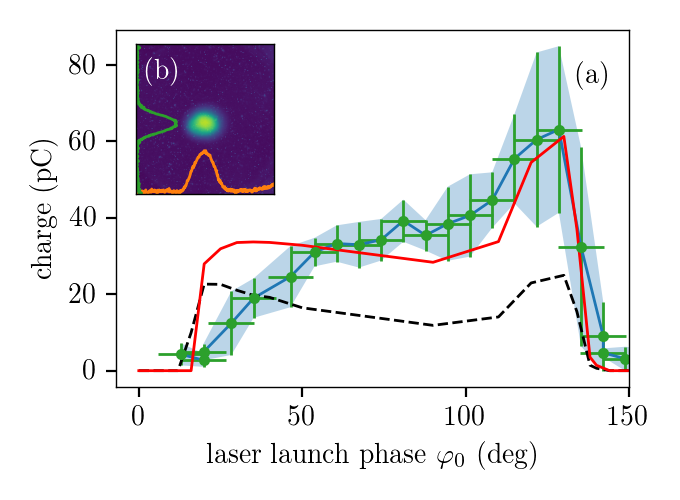}
\caption{Measured bunch charge as a function of laser-launch phase (circles with shaded area giving the uncertainty) compared with {\sc astra} simulations with (solid line, Schottky-strength parameter set to $\xi=3.5\times 10^{-3}$) and without (dash line, $\xi=0$) including the Schottky effect. The inset (b) gives the transverse beam distribution on a $10\times 10$~mm$^2$ area at X1 for $\varphi_0\simeq 95^{\circ}$ with associated projections (solid lines).}
   \label{fig:phasescan}
\end{figure}

After conditioning, the UV laser pulse was injected into the XRF gun. Coarse and fine timing scans were performed to find the optimal injection time of the laser pulse into the RF pulse. The coarse scans were performed by adjusting the optical delay of the incoming laser to vary its launch phase $\varphi_0$ over several RF cycles. During these ``phase scans" the field experienced during emission varies as $E_a=E_0\sin\varphi_0$ and the photoemitted charge is recorded leading to the localization of several RF buckets. Subsequently, fine scans were performed over the emission period $\varphi_0\in [0, 360]^{\circ}$ as shown in Fig.~\ref{fig:phasescan}(a) and confirmed the photoemission of bunches with charges up to $q=60\pm 25$~\si{\pico\coulomb}. The evolution of the emitted charge is also characteristic of Schottky-assisted photo-emission resulting from the lowering of the potential barrier in presence of applied high electric fields at the photocathode surface~\cite{schottky-1923-a}. The {\sc astra} beam-dynamics program~\cite{astra} was employed to simulate the phase scan. The simulation without accounting for the Schottky effect shows the phase-scan follows a plateau-shaped distribution with a slight dip for $\varphi_0\in[50,100]$ attributed to particle loss on the iris aperture. However, including the Schottky effect using a model where the charge of the $N$ macroparticles representing the bunch in {\sc astra} scales as $q_m =q/N+ \xi \sqrt{E_0\sin\varphi_m}$ (where $\varphi_m\equiv \varphi_0 +2\pi f t$ is the emission phase of the $m$-th macroparticle and $\xi$ controls the strength of the Schottky effect), produces a simulated phase-scan distribution with similar features as on the measured scan; see Fig.~\ref{fig:phasescan}(a). The transverse beam density measured at X1 appears in Fig.~\ref{fig:phasescan}(b).
\begin{figure}[bbbbbbb!!!!!]
   \centering
   \includegraphics[width=0.9\columnwidth]{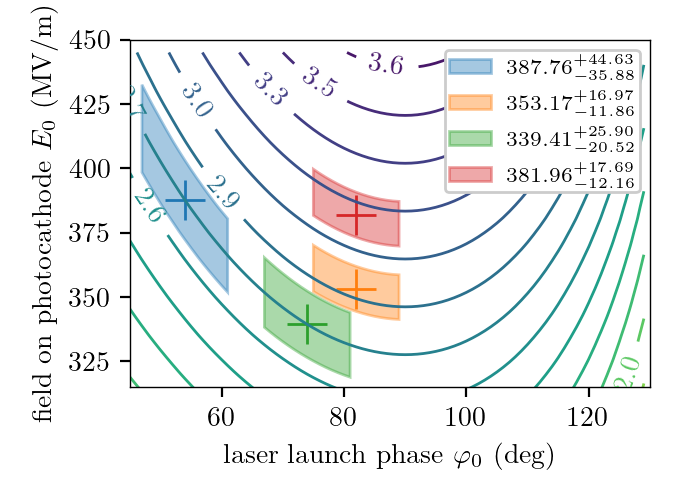}
\caption{Simulated kinetic-energy isoclines as function of RF-gun operating conditions $K(E_0,\varphi_0)$ and retrieved operating points (``$+$" symbols) for four measurements. The labeled isoclines gives the kinetic energy in \si{\mega\electronvolt} units. The shaded areas represent the uncertainty on the measured $\varphi_0$ and inferred $E_0$ values (reported in the legend in MV/m units).}
   \label{fig:errorregion}
\end{figure}

The energy measured downstream of the gun is a nonlinear function of the electric field experienced by the electron during the emission $E_a=E_0\sin \varphi_0$. More specifically, the beam's final kinetic energy is a bivariate function $K(E_0,\varphi_0)$ obtained by solving a set of coupled ordinary differential equations~\cite{kim-1989-a,floettmann-2015-a}. We used  \textsc{astra} to numerically compute  the function $K(E_0,\varphi_0)$ over a $28\times 86$ two-dimensional grid for  $E_0\in[315,450]~\si{\mega\volt/\metre}$ and $\varphi_0\in[45^{\circ},130^{\circ}]$. A spline-interpolated function $K(E_0,\varphi_0)$ was then used to calculate the value of $E_0$ given the measured kinetic energy and operating phase $\varphi_0$. The contour plot of $K(E_0,\varphi_0)$ and inferred fields values appear in Fig.~\ref{fig:errorregion}. The area enclosed by two kinetic-energy isoclines and two vertical lines of operating phase, obtained from measurement uncertainties, provide the error region of the calculated field with its upper and lower error bounds. The result of such an analysis confirms that the maximum peak field attained during our experiment was  $E_0=387.76^{+44.63}_{-35.88}~\si{\mega\volt/\metre}$ corresponding to a surface field at the iris of $1.55E_0\simeq 601.03^{+69.18}_{-51.61}$~\si{\mega\volt/\metre}. These field strengths were consistently reached over extended periods during the three-week experimental run.

\begin{figure}[tttttt!!!!]
   \centering
   \includegraphics[width=0.95\columnwidth]{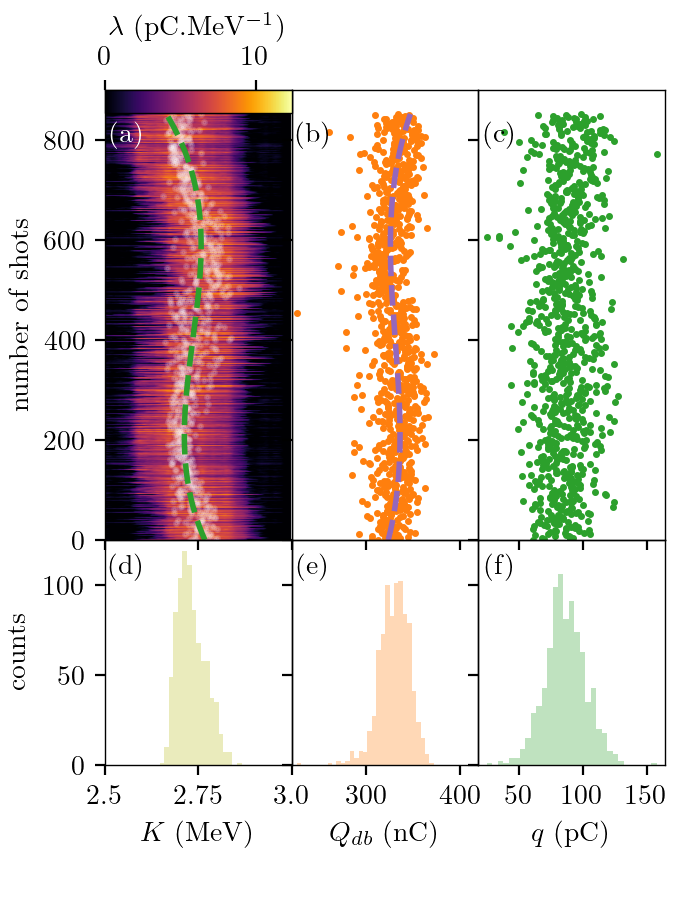}
\caption{Waterfall plot of measured energy spectra density $\lambda$ over 853 shots (a) with associated drive-beam charge transmitted through the PETS (b), and corresponding bunch charge photoemitted from the XRF gun (c). Plots (d-f) are the associated histograms. The dash trace in (a) is a spline interpolation of the slow drift which is superimposed as a dashed curve in (b) after applying an affine transformation. }
   \label{fig:shots}
\end{figure}

Figure~\ref{fig:shots}(a) displays a waterfall plot of stacked energy spectra for successive shots acquired over a 15-mins period. The data reveal the kinetic energy jitter over the period to be $\frac{\Delta K}{K} \simeq 1.4$\% (rms) and to consist of a shot-to-shot jitter superimposed with a long-term ($\sim 5$~mins) drift. The latter slow oscillation appears to be partially correlated to the drive-beam charge fluctuation [see Fig.~\ref{fig:shots}(a) and (b)]. During our experiment the drive-beam charge relative fluctuation was $\frac{\Delta Q_{db}}{Q_{db}} \simeq 4.9$\% (rms) and the bunch charge photoemitted from the XRF gun was $q=86.2\pm 16.1$~pC; see Fig.~\ref{fig:shots}(e,f). The XRF-gun charge jitter is dominated by phase and amplitude jitters due to the drive-beam jitter and drift which impacts $Q_{db}$; see Fig.~\ref{fig:shots}(d). The overall stability of the XRF gun is limited by the current experimental configuration. Removing long-term drift indicates that the $Q_{bd}$ has a  shot-to-shot jitter of 4.7\% comparable to the UV-laser energy jitter on the drive-beam photocathode. \\

In conclusion, we demonstrated the generation of relativistic ($\sim 3$-MeV) electron bunches from an ultra-high gradient X-band RF gun powered by nanoseconds RF pulses. The gun supports photocathode fields $E_0\sim 400$~MV/m $-$ i.e., $\sim 3$ times larger than reliably attained in previous setups $-$ without any observable breakdown or detectable dark current. Such an accomplishment confirms that powering accelerating structures with short RF pulses provides a viable path to establishing GV/m peak electric fields for generating bright ultra-relativistic electron bunches with applications to advanced accelerators and compact light sources. \\ 

We thank V. Dolgashev (SLAC National Accelerator Laboratory) for his generous support of X-band RF components.This work was supported by the U.S. DOE, under awards No. DE-SC0018656 and DE-SC0022010 to Northern Illinois University, DOE SBIR grant No DE-SC0018709 to Euclid Techlabs LLC. The research uses resources at Argonne National Laboratory funded by the U.S. DOE under contract No. DE-AC02-06CH11357.  

\bibliography{apssamp}

\appendix

\end{document}